\documentstyle[emulateapj,10pt,apjfonts,psfig]{article}

\newcommand{\ks}{km s$^{-1}$~}
\newcommand{\kms}{km s$^{-1}$~}

\def\HI{H\,{\sc i}}

\def\farcm{\hbox{$.\mkern-4mu^\prime$}}
\def\etal{et~al.\ }

\lefthead{DUBNER ET AL}
\righthead{THE INTERSTELLAR MEDIUM AROUND SNR~G320.4--1.2}

\begin{document}

\title{The Interstellar Medium around the Supernova~Remnant~G320.4--1.2 }
\submitted{Accepted to The Astronomical Journal}

\author{G. M. Dubner\altaffilmark{1,2}, 
B. M. Gaensler\altaffilmark{3,4,5}, 
E. B. Giacani\altaffilmark{1,2}, 
W. M. Goss\altaffilmark{6} and
A. J. Green\altaffilmark{7}}

\altaffiltext{1}{Instituto de Astronom\'\i a y F\'\i sica del Espacio
(CONICET, UBA), C.C.67, 1428 Buenos Aires, Argentina;
gdubner@iafe.uba.ar}
\altaffiltext{2}{Member of the Carrera del Investigador
Cient\'\i fico of CONICET, Argentina}
\altaffiltext{3}{Center for Space Research, Massachusetts 
Institute of Technology, Cambridge, MA 02139}
\altaffiltext{4}{Hubble Fellow}
\altaffiltext{5}{Current address:
Harvard-Smithonian Center for Astrophysics,
60 Garden Street, Cambridge, MA 02138}
\altaffiltext{6}{National  Radio  Astronomy  Observatory,
P.O. Box 0, Socorro, NM 87801}
\altaffiltext{7}{School of Physics, University of Sydney, NSW 2006, Australia} 

\begin{abstract}

Using the Australia Telescope Compact Array, we have carried out a
survey of the \HI\
emission in the direction of the ``barrel-shaped'' supernova remnant (SNR) G320.4--1.2
(MSH 15--5{\em 2}) and its associated young pulsar B1509--58. The angular 
resolution of the data is $4\farcm0 \times 2\farcm7$, and the rms noise of the order of 30 mJy/beam ($\sim 0.5$ K).  
 The \HI\ observations
indicate that the  N-NW radio limb has encountered a dense \HI\ filament 
(density $\sim $ 12 ~cm$^{-3}$) at the same LSR velocity
than that of the SNR (V$_{\rm LSR} \sim -68$ \kms). This \HI\ concentration 
would be  responsible for the
flattened shape of the NW lobe of G320.4-1.2, and for the formation of the 
radio/optical/X--ray nebula RCW 89. The emission associated with the 
bright knots in the interior of RCW 89 can be explained as arising from the 
interaction between the collimated relativistic outflow from the pulsar and 
the denser part of this \HI\ filament (density $\sim $ 15 ~cm$^{-3}$). 
The S-SE half of the SNR, on 
the other hand, seems to have rapidly  expanded   across  a 
lower density enviroment  
 (density $\sim 0.4$ cm$^{-3}$).  
The \HI\ data also reveal an unusual \HI\ feature aligned 
 with a collimated outflow generated by
the pulsar, suggestive of association with the SNR. The anomalous kinematical 
velocity of this feature (V$_{\rm LSR} \sim 15$ \kms), however, is difficult 
to explain. 

\end{abstract} 

\keywords {
ISM: individual (G320.4--1.2, RCW~89, MSH15-5\it2) ---
ISM: structure ---
pulsars: individual (B1509--58) ---
radio lines: ISM ---
supernova remnants
}

\section{Introduction}

The radio morphologies of supernova remnants (SNRs) are usually
approximated by limb-brightened spherical shells, in which the shock
front accelerates electrons  to synchrotron-emitting energies.
However, observations at increasingly higher resolution and sensitivity
have demonstrated that actual SNR morphologies are much more complex,
exhibiting ``break-out'', ``barrel''
and helical patterns (\cite{man87};
\cite{dgg+96}; \cite{wg96}; \cite{gae98}; \cite{dhgm98}).
Such a diversity reflects not only variations in the
properties of the progenitor star and of the explosion itself, but also
echoes the physical conditions of circumstellar and interstellar gas
and magnetic fields.  The interpretation of SNR
morphologies has the additional complication that the images 
observed are a two-dimensional projection  of a three-dimensional
object.  Radio spectroscopy of the interstellar medium (ISM) 
provides a valuable tool 
to unravel these 3D effects using kinematic signatures.
As part of an on-going effort to study the interaction
of SNRs with their environment (\cite{dgg+98}; \cite{dhgm98}; \cite{dgr+99};
\cite{gdg+00}), we 
present a study of the distribution and kinematics
of the neutral hydrogen in the vicinity of SNR~G320.4--1.2 
(MSH 15--5{\em 2}).

SNR G320.4--1.2 has
an unusual radio morphology, consisting of 
two loosely connected emission regions, 
one to the N-NW and the other to the S-SE.
Figure~\ref{fig1} shows the radio and X-ray emission from the SNR.
The young pulsar PSR~B1509-58 lies between the
two main radio components of the SNR; the pulsar's 
position as measured by Gaensler \etal\ (1999\nocite{gbm+98}), hereafter G99,
is marked in Figure~\ref{fig1} with a cross.
The optical/radio/X-ray nebula RCW~89 is  located on the NW extreme of  
G320.4--1.2.  

Like many other SNRs, G320.4--1.2 has
a ``bilateral'' or ``barrel'' morphology (\cite{kc87}; \cite{gae98}).
Many different mechanisms have been proposed
to account for this morphology, relating to both the
progenitor star and supernova explosion (``intrinsic effects'')
(\cite{kc87}; \cite{man87}; \cite{bd94}; \cite{wwps96})
and to subsequent interaction with the SNR environment (``extrinsic effects'')
(\cite{bls90b}; \cite{fr90}). Most recently,
Gaensler (1998\nocite{gae98}) has argued that
the bilateral appearance arises when a SNR 
expands into an elongated
low-density cavity; the radio emission then originates at the sites where the
shock encounters the walls of the cavity.

This peculiar SNR and its energetic young pulsar 
have been extensively observed throughout
the electromagnetic spectrum (see G99
for a summary of previous observations). In particular, recent
X-ray and radio  data provide convincing evidence
that the SNR and pulsar resulted from
a single supernova explosion which occurred approximately
1700~yrs ago (\cite{tkyb96}; \cite{bb97}; G99\nocite{gbm+98};
\cite{gak+01}).
Specifically, G99 argue that the
entire complex can be interpreted as the result of
a low-mass or high-energy explosion 
occurring near one edge of an elongated low density
(n $\sim 0.01$~cm$^{-3}$) cavity. To the NW
the SNR has expanded into dense material (n $\sim 1-5$~cm$^{-3}$),
while to the SE the remnant has expanded rapidly across the
low density cavity. 
Meanwhile, the pulsar appears to be
generating twin collimated outflows,
the northern of which is interacting with the SNR
to produce a collection of compact radio/X-ray
knots embedded in RCW~89. A distance of (5.2 $\pm$ 1.4) kpc was derived
for G320.1--1.2 by G99, based on \HI\ absorption measurements, in
agreement with the results of Caswell et al. (1975).

Within this complex scenario, it is clear that the interpretation of 
the existing body of observational data can greatly
benefit from  an accurate
knowledge of the distribution and kinematics of the surrounding  
neutral hydrogen emission.
 Here we report on 
an \HI\ survey of SNR~G320.4--1.2 and its environment, carried out
with  the Australia Telescope
Compact Array. 

\section{Observations and Data Reduction}
\label{sec_obs}

Interferometric \HI\ observations towards
G320.4--1.2 were carried out using the Australia
Telescope Compact Array (ATCA; \cite{fbw92}),
a six-element synthesis telescope
located near  Narrabri, NSW, Australia,
during a session of 13 hours on 1998 October 13. In order to 
achieve  optimal $u-v$ coverage in the radial direction, the array was
used  in the non-standard 210-m configuration, in
which five antennas are positioned closely together
so as to provide 10 baselines ranging from 31 to 214~m.
The sixth (fixed) antenna is separated from
the other five elements by 5--6~km, and was not used in our analysis.

A region surrounding the SNR was surveyed in
a mosaic of 19 pointings, following a hexagonal grid to  cover
six square degrees . The separation between grid-points of
about $16'$ satisfies the Nyquist sampling criterion
for the primary beam width of $33'$ (FWHM).
Observations were centered
at a frequency of 1420~MHz, using
1024 channels over a total bandwidth of 4~MHz. The absolute flux density
scale was determined by assuming a flux density
for PKS~B1934--638 of 14.85~Jy at 1.42~GHz. Observations
of PKS~B1934--638 were also used to determine the spectral response across
the observing band. Time variability
in the antenna gains was calibrated using regular observations
of PKS~B1540--828. After editing and calibration,
continuum emission
was subtracted from the data in the $u-v$ plane
using the task {\tt UVLSF}\ in {\tt AIPS} (\cite{vc90}).
The data were then transferred to the {\tt MIRIAD}\ package
(\cite{sk98}),
where all pointings were imaged simultaneously to form 390 mosaicked planes,
covering LSR velocities between --150 and +171~\kms\
at a velocity resolution of 1~\kms (channel separation of 0.82~\kms). This cube
was then deconvolved using a mosaicked maximum-entropy
technique (\cite{ssb96}), and
then convolved with a gaussian restoring beam.

Any analysis of \HI\ emission requires measurements at all spatial
scales down to the resolution limit. However, our ATCA observations are
not sensitive to structures larger than $15'-20'$.  To recover this
short-spacing data, we have utilized \HI\ observations made using the
Parkes 64-m radio-telescope, carried out as part of the Southern
Galactic Plane Survey (SGPS; \cite{mgd+01}) with an
angular resolution of $14'$. The Parkes data were multiplied by a
factor of 0.9 to match the flux calibration applied to the ATCA
data; interferometric and single-dish data were then appropriately
weighted according to
their respective primary beam shapes and 
combined using the task {\tt IMMERGE}\ in {\tt
MIRIAD} (\cite{sta99}; \cite{sk98}).
The final combined cube has units of Jy~beam$^{-1}$.
To convert to units of brightness temperature, these data must 
be multiplied by a factor of 16.3~K~Jy$^{-1}$~beam.  
The angular resolution of the data is $4\farcm0 \times 2\farcm7$
(Position Angle $55^\circ$), and the 
rms in line-free channels is $\sim30$~mJy~beam$^{-1}$
($\sim 0.5$~K).  

\section{Results}
\label{sec_results}

\subsection{General \HI\ distribution}
\label{sec_results_general}

Figure~\ref{fig2} illustrates the general 
properties of \HI\ emission in the direction of G320.4--1.2.
The top panel shows an average \HI\ profile of the observed region
taken from the Parkes SGPS data, while
the bottom panel displays the circular rotation model towards $l= 320^\circ$, 
$b= -1^\circ$, assuming
$R_0=8.5$~kpc and $\Theta_0=220$~\kms\ and using
the rotation curve of Fich, Blitz \& Stark (1989\nocite{fbs89}).

The neutral hydrogen emission in this direction has
a complex structure, and is
spread over more than 200~\kms. The five main peaks of this spectrum
can be interpreted as emission from gas related to the Sagitarius-Carina
galactic arm (that the line of sight crosses at both $V_{\rm LSR}\sim -50$
and $\sim +80$~\kms), the Scutum -Crux arm (at velocities of 
 $V_{\rm LSR}\sim -20$ and + 40 ~\kms), and to the local gas contribution
near 0 \kms. The SNR would be located on the far border of Carina arm
(Georgelin et al. 1987).

We investigate the \HI\ emission distribution in the entire
observed velocity interval, looking for morphological and/or
kinematical evidences of interaction of G320.4--1.2 and its different
components with the surrounding \HI\ gas.
Figure~\ref{fig3} provides an overview of the distribution
of atomic hydrogen in the surveyed
region. It displays the \HI\ emission  between $\sim -81$~\kms\ 
and $\sim +92$~\ks, 
the interval in which significant \HI\
emission is found. Each image
is the average of six consecutive spectral channels,  spanning 5 \ks.
However the analysis is carried out with the highest possible velocity
resolution. 
The field shown is slightly reduced in size with
respect to the observed field to avoid edge effects. The velocity shown
in the lower left  
corner of each image corresponds to the velocity of
the first integrated channel. Superimposed on the \HI\
images are a few representative contours of the radio continuum emission of
G320.4--1.2 at 1.4 GHz as taken from G99, at a spatial
resolution of $24''\times21''$.

 From $\sim -76$ \ks to $\sim -66$ \ks, the most conspicuous feature in
\HI\ is an elongated filament seen partially overlapping the bright NW part
of the SNR. This filament runs diagonally
from NE to SW, in a direction parallel to the Galactic plane. 
This filament,
is discussed in detail in the next section.

 From $\sim$ --60 to $\sim$ --40 \ks, a strong gradient in the gas 
density (assumed to be optically
thin) between the NW half of the field (closer to the Galactic plane)
and the SE corner (away from the Galactic plane) is evident. 
This density gradient persists up
to $V \sim -31$~\ks, although with less contrast than at more
negative velocities.  Significant absorption is evident
against bright continuum emission from RCW~89 in this velocity range
(See also G99).

 From $\sim -26$ to $\sim 0$ \ks, the \HI\ emission appears quite 
uniformly distributed, with a few brighter concentrations near the NW corner.
Strong absorption against RCW~89 is observed at these velocities.
In the most severe cases, this absorption is surrounded
by residual sidelobes which were not completely removed
in the deconvolution process.

Positive velocities along this line-of-sight
correspond to 
distant gas, located beyond $\sim12$~kpc (see Fig~\ref{fig2}). We have
examined the \HI\ emission at positive velocities,
bearing in mind that these data can also
include contributions from closer gas which might
be associated with G320.4--1.2 but which has been kinematically disturbed.

In the velocity range between --1.5 and 8~\kms\ 
(i.e.\ in the images labeled --1.5~\kms\ and +3.5 \kms in Fig. 3),  
two parallel bands of \HI\ emission  cross over the field from NE to SW. The
\HI\ concentration that runs close to the SE lobe of G320.4--1.2, shows 
good morphological correspondence with the radio continuum emission of the
SNR. This \HI\ filament, as seen  in projection,  surrounds the SE
lobe of the SNR  along the eastern and southern borders. Such correspondence is
even closer when comparing the \HI\ distribution with the X-ray
emission. However, since at this kinematical velocity there is no
 way to disentangle the
contribution of gas likely to be  associated with G320.4--1.2 
from the intense local gas emission, we do not
consider this structure as a firm candidate for  \HI\--SNR association. 
 
From  $\sim$ +13 to $\sim$ +22 \ks, there is an unusual, bright filament
oriented almost perpendicular to the Galactic plane,
breaking the plane-parallel symmetry observed otherwise.
This filament overlaps both the brightest region of the SNR and the pulsar 
itself, and  will be discussed further in the next section.

At +33 \ks and +38~\ks, a long, moderately bright \HI\ filament is
observed 
in apparent contact with the NW half of the SNR. However, since most of
the \HI\   in this direction is expected to have
this plane-parallel distribution,
we conclude that this feature is most probably
distant background material, located 
14 kpc away. The remaining
high-velocity images show no features which can be feasibly
associated with G320.4--1.2.

In order to determine the systemic velocity of G320.4-1.2, we can 
use Fig.3 to analize the characteristics of the HI near the SNR. In
agreement  
with previous HI absorption studies ( Caswell et al. 1975, G99), the images 
in Fig. 3 show that HI absorption is present towards RCW 89 in the 
velocity range  between --61~\kms and --51~\kms. Thus, the kinematical velocity corresponding to
G320.4-1.2 must be more negative than --61 \kms. In addition, \HI\ is
observed  in emission near the northern portion of the remnant up
 to $V_{\rm LSR}\simeq -76$~\kms. We therefore conclude that 
the systemic velocity of G320.4-1.2 lies in the range 
--76 to --61 \kms. Adopting a systemic velocity of 
$V_{\rm LSR}\sim -68$ \kms, the kinematical distance to the 
source turns out to be  about 5 kpc, in good agreement with a previous estimate by G99. 
 
\subsection {Particular \HI\ Features}

In this section we focus on specific \HI\ features,
searching for favorable  \HI\--SNR associations. 

{\em 1. The  northern filament in the interval 
$-76 \la V_{\rm LSR} \la -66$~\kms:} This velocity interval 
 corresponds to the systemic velocity of G320.4-1.2. 
Figure~\ref{fig4} shows in greyscale  the \HI\ distribution obtained by
integrating between velocities
of --76 and --66 \ks.  The 1.4 GHz 
 radio continuum emission from G320.4-1.2 (as taken from G99) is plotted 
in contours. 
This extended \HI\ filament to the NW may represent 
the dense external wall previously suggested by G99 to be slowing down the 
expansion of G320.4-1.2 in this direction. This feature may also be responsible for 
the flattened shape of the NW lobe of the bilateral SNR. In particular 
 this filament has a peak
in HI  emission ( brightness temperature 80 K at v=--70 \kms)  
exactly overlapping  RCW89.  
 We propose
that this \HI\ cloud (about 7' in diameter) associated with RCW89, 
is the dense material with which 
 the SNR has collided and with which the outflow
from the pulsar is interacting (\cite{tkyb96}; G99).
According to this interpretation, the
bright nebula RCW89 (or, more precisely,  the bright radio/X-ray knots
in its interior) has
formed  at the site of interaction between the 
collimated relativistic outflow from the pulsar and the dense  \HI\ cloud.
The optical emission associated with RCW89 can be explained as arising from
 the ionized outer layers of this high density
material, as  first suggested by Manchester (1987\nocite{man87}). 

The total \HI\ mass of the extended filament has been estimated in M$\sim$ 4000
M$_\odot$, where an appropriate background correction has been applied.
An average atomic density n$\sim$ 12 cm$^{-3}$ has been calculated for
this elongated \HI\ feature by assuming that the dimension along the
line of sight equals the size of the minor axis of the filament. In
particular, for the denser concentration of this feature, associated
with RCW 89, a total mass M$\sim$ 1700 M$_\odot$ and a volume density
n$\sim$ 15 cm$^{-3}$, have been derived (in this case a spherical
geometry was assumed for the \HI\ feature).

As another evidence of the interaction of G320.4--1.2 with its
surroundings, we have searched for accelerated shocked \HI\ gas all
along the northern \HI\ filament, and especially in the direction of
RCW 89, where the major interaction SNR--ISM occurs. We examined several
\HI\ spectra, as well as RA vs V$_{LSR}$ plots (traced at constant
declination), looking for high-velocity features, as observed for
example in IC 443 (DeNoyer 1977). Within the moderate angular
resolution of the present data, no evidences of accelerated \HI\ gas
were found.

We have also investigated the characteristics of the interstellar gas
where the SE lobe of G320.4-1.2  has expanded. To carry out this calculation we
have considered a sample volume in a location relatively free of \HI\
concentrations around V$_{LSR} \sim -68$ \kms, the systemic velocity of
G320.4--1.2. The obtained volume density is n$\sim 0.4$ cm$^{-3}$.

The observed contrast in density between the \HI\ near the 
NW lobe  and the ambient \HI\ in
the interior of the SNR,  is
consistent with the model of G99 in which the  NW half of the SNR has 
encountered the edge of a cavity, while the SE half has expanded 
relatively unimpedded into a lower density environment. 

{\em 2. The bright filament perpendicular to the Galactic plane in the
interval $+10 \la V_{\rm LSR} \la +20$~\kms:}
Figure~\ref{fig5} displays in detail this \HI\
filament (thick black contours), overlaid on radio continuum
(greyscale) and X-ray (light grey contours) emission from the SNR. 
The cross
indicates the position of PSR~B1509--58. 

 From a morphological point of
view, there is a striking alignment of this \HI\ filament with both the
NW/SE axis of the X-ray nebula and the weak radio features
extending south of RCW~89. {\em ROSAT} and {\em Chandra} data
demonstrate 
that the elongated X-ray nebula represents
opposed collimated outflows directed (in projection) along this NW/SE axis
(\cite{bb97}; \cite{gak+01}).
G99 argue that the radio emission in this
region corresponds to a cylindrical sheath surrounding
the pulsar outflow; 
it has been shown that the magnetic field in the region is
well-ordered and oriented in a direction parallel to the axis of the
proposed outflow (\cite{mch93}; G99\nocite{gbm+98}).
It is thus conceivable that a pre-existing column 
of gas with an orientation similar to that of the outflow from the
pulsar may have provided the conditions necessary to form a
cylinder and thus generate the observed characteristics. 
For instance, the magnetic
field could be frozen to the column of cold \HI\, thus explaining its
order and orientation. This \HI\ feature may  suggest an association
with G320.4--1.2, based solely on the morphology. However,
the high difference between the systemic velocity of the SNR and the
\HI\ velocity, remains unexplained. For this feature a total \HI\ mass
M$\sim 2700$ M$_\odot$ and an average atomic density n$\sim 10$
cm$^{-3}$ have been estimated.

\section{ Conclusions}

We have conducted a study of the distribution and kinematics of the 
interstellar neutral hydrogen in a field around the bilateral SNR~G320.4--1.2.
This SNR comprises
the bright optical/X-ray/radio nebula RCW~89  and the energetic
young pulsar B1509--58.

We find several features in \HI\ which suggest that the radio
morphology of G320.4--1.2 is a result of the distribution of the 
surrounding ISM, 
combined with the action of the pulsar's relativistic outflows.

The N-NW bright radio limb of G320.4-1.2 
shows the effects of interaction with an extended \HI\ filament 
 with  density of $\sim$ 12  cm$^{-3}$ present at the same velocity as that of the SNR ($V_{\rm
LSR}\sim -68$~\kms). In addition, the interaction of the  relativistic
outflow from PKS B1509-58 with the denser part of this filament
(n$\sim$ 15 cm$^{-3}$) may have been responsible for the formation 
of the bright  nebula RCW89 located on the NW
extreme of G320.4-1.2. The opposite radio limb of the
SNR apparently expanded into a lower density environment (n $\sim$ 0.4 
cm$^{-3}$) with little distortion from a   semicircular shape. 

The nature of ``barrel''--shaped SNRs can be explained assuming that the
SNR expands into a low density cavity with walls almost parallel, and
the bright radio limbs originate where the expanding shock encounters
the walls (Gaensler 1998). In the case of G320.4--1.2, the present \HI\
study has confirmed the existence of the \HI\ wall associated with the
N--NW lobe. However, the observations are not conclusive with respect
to the S--SE \HI\ counterpart. We have identified \HI\ structures with
the location and morphology suggestive of an \HI\--SNR association in this 
direction.  
However, the fact that they are detected near V$_{LSR}\sim +3.5$ \kms,
indistinguishable in practice  
from local gas contribution, makes the association uncertain.

Based on morphological coincidences, we suggest
a physical association between the SNR and an
 elongated \HI\ feature with 
V$_{LSR}\sim +15$ \kms, oriented perpendicular to
the Galactic plane. This \HI\ structure 
is strikingly aligned with the projected axis inferred for the
collimated outflows generated by PSR~B1509--58. It is difficult
to explain the anomalous kinematical velocity of this feature;
we suggest that this gas is indeed at the same
distance as is G320.4--1.2, but suffers from kinematical distortions. Such
perturbations in the velocity field are not unexpected in this region,
rich in massive stars with powerful stellar winds that have stirred
the interstellar material for thousands of years (Georgelin et al.
1987; Lortet, Georgelin \& Georgelin 1987).

\acknowledgments

We thank the staff of the ATCA for technical support during the
observations, and Naomi McClure-Griffiths for providing us with Parkes
data from the Southern Galactic Plane Survey.  
  This research was partially
funded through a Cooperative Science Program between
CONICET (Argentina) and the National Science Foundation (USA) and 
through CONICET grant 4203/96.
  B.M.G.
acknowledges the support of NASA through Hubble Fellowship grant
HST-HF-01107.01-A awarded by the Space Telescope Science Institute,
which is operated by the Association of Universities for Research in
Astronomy, Inc., for NASA under contract NAS 5--26555. The Australia
Telescope is funded by the Commonwealth of Australia for operation as a
National Facility, managed by CSIRO. The National
Radio Astronomy Observatory is a facility of the National Science
Foundation  operated under cooperative agreement by Associated
Universities, Inc.

\bibliographystyle{apj1}

\begin{thebibliography}{}

\bibitem[Bisnovatyi-Kogan, Lozinskaya, \& Silich 1990]{bls90b}
Bisnovatyi-Kogan, G.~S., Lozinskaya, T.~A., \& Silich, S.~A. 1990, { \apss},
  {\rm 166}, 277.

\bibitem[Brazier \& Becker 1997]{bb97}
Brazier, K. T.~S. \& Becker, W. 1997, { MNRAS}, {\rm 284}, 335.

\bibitem[Brighenti \& D'Ercole 1994]{bd94}
Brighenti, F. \& D'Ercole, A. 1994, { MNRAS}, {\rm 270}, 65.

\bibitem[Caswell \etal  1975]{cmr+75}
Caswell, J.~L., Murray, J.~D., Roger, R.~S., Cole, D.~J., \& Cooke, D.~J. 1975,
  { A\&A}, {\rm 45}, 239.


\bibitem[DeNoyer 1977]{dn77}
DeNoyer, L.K., 1977, {ApJ}, {\rm 212}, 416.

\bibitem[Dubner \etal  1999]{dgr+99}
Dubner, G., Giacani, E., Reynoso, E., Goss, W.~M., Roth, M., \& Green, A. 1999,
  { AJ}, {\rm 118}, 930.

\bibitem[Dubner \etal  1996]{dgg+96}
Dubner, G.~M., Giacani, E.~B., Goss, W.~M., Moffett, D.~A., \& Holdaway, M.
  1996, { AJ}, {\rm 111}, 1304.

\bibitem[Dubner \etal  1998a]{dgg+98}
Dubner, G.~M., Green, A.~J., Goss, W.~M., Bock, D., C.-J., \& Giacani, E.
  1998a, { AJ}, {\rm 116}, 813.

\bibitem[Dubner \etal  1998b]{dhgm98}
Dubner, G.~M., Holdaway, M., Goss, W.~M., \& Mirabel, I.~F. 1998b, { AJ}, {\rm
  116}, 1842.

\bibitem[Fich, Blitz, \& Stark 1989]{fbs89}
Fich, M., Blitz, L., \& Stark, A.~A. 1989, { ApJ}, {\rm 342}, 272.

\bibitem[Frater, Brooks, \& Whiteoak 1992]{fbw92}
Frater, R.~H., Brooks, J.~W., \& Whiteoak, J.~B. 1992, { J. Electr. Electron.
  Eng. Aust.}, {\rm 12}, 103.

\bibitem[Fulbright \& Reynolds 1990]{fr90}
Fulbright, M.~S. \& Reynolds, S.~P. 1990, { ApJ}, {\rm 357}, 591.

\bibitem[Gaensler 1998]{gae98}
Gaensler, B.~M. 1998, { ApJ}, {\rm 493}, 781.

\bibitem[Gaensler \etal  2001]{gak+01}
Gaensler, B.~M., Arons, J., Kaspi, V.~M., Pivovaroff, M.~J., Kawai, N., \&
  Tamura, K. 2001, { ApJ}, 
\newblock submitted.

\bibitem[Gaensler \etal  1999]{gbm+98}
Gaensler, B.~M., Brazier, K. T.~S., Manchester, R.~N., Johnston, S., \& Green,
  A.~J. 1999, { MNRAS}, {\rm 305}, 724 (G99).

\bibitem[Georgelin \etal  1987]{gbg+87}
Georgelin, Y.~M., Boulesteix, J., Georgelin, Y.~P., Laval, A., \& Marcelin, M.
  1987, { A\&A}, {\rm 174}, 257.

\bibitem[Giacani \etal  2000]{gdg+00}
Giacani, E.~B., Dubner, G.~M., Green, A.~J., Goss, W.~M., \& Gaensler, B.~M.
  2000, { AJ}, {\rm 119}, 281.

\bibitem[Kesteven \& Caswell 1987]{kc87}
Kesteven, M.~J. \& Caswell, J.~L. 1987, { A\&A}, {\rm 183}, 118.

\bibitem[Lortet, Georgelin, \& Georgelin 1987]{lgg87}
Lortet, M.~C., Georgelin, Y.~P., \& Georgelin, Y.~M. 1987, { A\&A}, {\rm 180},
  65.

\bibitem[Manchester 1987]{man87}
Manchester, R.~N. 1987, { A\&A}, {\rm 171}, 205.

\bibitem[McClure-Griffiths \etal  2001]{mgd+01}
McClure-Griffiths, N.~M., Green, A.~J., Dickey, J.~M., Gaensler, B.~M., Green,
  A.~J., Haynes, R.~F., \& Wieringa, M.~H. 2001, { ApJ}, {\rm 551}, 394.

\bibitem[Milne, Caswell, \& Haynes 1993]{mch93}
Milne, D.~K., Caswell, J.~L., \& Haynes, R.~F. 1993, { MNRAS}, {\rm 264}, 853.

\bibitem[Sault \& Killeen 1999]{sk98}
Sault, R.~J. \& Killeen, N. E.~B. 1999, { The Miriad User's Guide}, (Sydney:
  Australia Telescope National Facility).
\newblock (http://www.atnf.csiro.au/computing/software/miriad/).

\bibitem[Sault, Staveley-Smith, \& Brouw 1996]{ssb96}
Sault, R.~J., Staveley-Smith, L., \& Brouw, W.~N. 1996, { A\&AS}, {\rm 120},
  375.

\bibitem[Stanimirovi\'{c} 1999]{sta99}
Stanimirovi\'{c}, S. 1999.
\newblock PhD thesis, University of Western Sydney.

\bibitem[Tamura \etal  1996]{tkyb96}
Tamura, K., Kawai, N., Yoshida, A., \& Brinkmann, W. 1996, { PASJ}, {\rm 48},
  L33.

\bibitem[Trussoni \etal  1996]{tmc+96}
Trussoni, E., Massaglia, S., Caucino, S., Brinkmann, W., \& Aschenbach, B.
  1996, { A\&A}, {\rm 306}, 581.

\bibitem[van Langevelde \& Cotton 1990]{vc90}
van Langevelde, H.~J. \& Cotton, W.~D. 1990, { A\&A}, {\rm 239}, L5.

\bibitem[Whiteoak \& Green 1996]{wg96}
Whiteoak, J. B.~Z. \& Green, A.~J. 1996, { A\&AS}, {\rm 118}, 329.
\newblock (http://www.physics.usyd.edu.au/astrop/wg96cat/).

\bibitem[Willingale \etal  1996]{wwps96}
Willingale, R., West, R.~G., Pye, J.~P., \& Stewart, G.~C. 1996, { MNRAS}, {\rm
  278}, 749.

\end{thebibliography}

\clearpage

\begin{figure*}[hbt]
%\centerline{\psfig{file=Dubner.fig1.ps,height=12cm}}
\caption{{\em Left:} Radio continuum image of the SNR G320.4-1.2 at
1.4~GHz, at a resolution of $24'' \times 21''$ (G99).
The greyscale varies linearly from 0 to +75 mJy~beam$^{-1}$, while
contours are at levels of 5, 10, 20, 30, 60, 90, 120 and
150~mJy~beam$^{-1}$.  A few Galactic coordinate lines are included for
reference.  {\em Right:} Greyscale and contour {\em ROSAT}\ PSPC X-ray
image of G320.4--1.2 (\cite{tmc+96}). The  contour levels (in
arbitrary units) are 0.5, 1, 1.5, 2, 4 6, 10, 15 and 20. The white cross
indicates the position of PSR~B1509--58 as determined from
interferometric measurements (G99).\label{fig1}}  
\end{figure*}

\begin{figure*}[hbt]
\centerline{\psfig{file=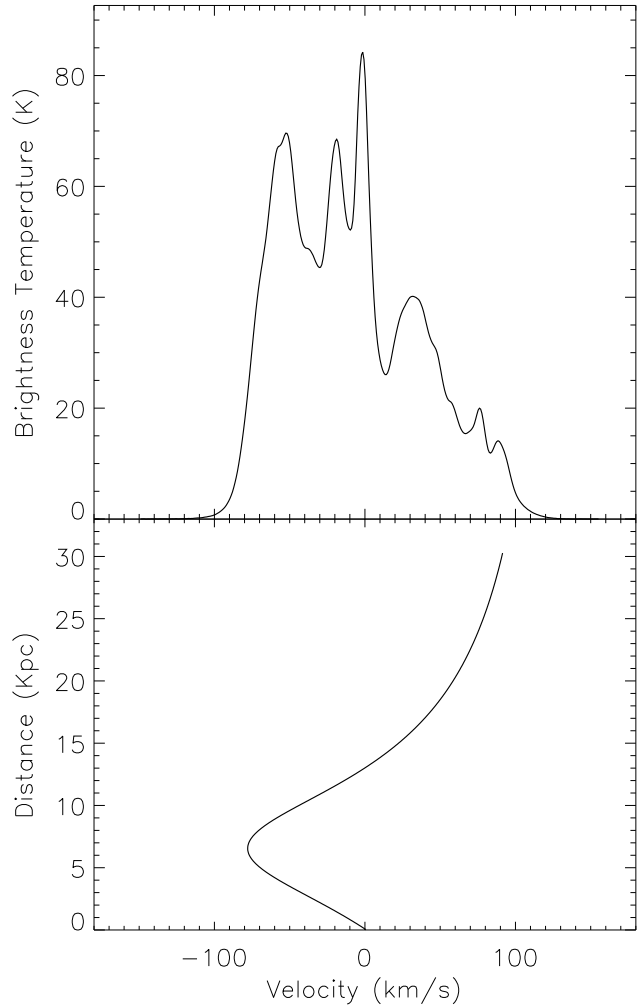,height=15cm}}
\caption{{\em Top:}  Average profile   of the \HI\ emission 
as taken from Parkes measurements  
towards SNR~G320.4--1.2, obtained as part of the Southern
Galactic Plane Survey.  {\em Bottom:} Galactic rotation curve for $l
= 320^\circ$, $b= -1^\circ$, according to the Galactic  circular rotation model
of Fich, Blitz \& Stark (1989\protect\nocite{fbs89}).\label{fig2}}
\end{figure*}

\begin{figure*}[hbt]
\caption{Images of the \HI\ emission between LSR velocities of $\sim
-81$ and $\sim +92$~\kms, each integrated over six consecutive channels
(approximately 5~\kms).  The velocity shown in the lower left corner of
each image corresponds to the first integrated channel; the grey scale
for the images varies between +0.08 and +6~Jy~beam$^{-1}$~\ks.  The
resolution of the data is $4\farcm0 \times 2\farcm7$ at a position
angle of $55^\circ$ (measured north through east).  Superimposed on the
\HI\ images are a few representative contours of the radio continuum
emission of G320.4-1.2 at 1.4~GHz (G99).\label{fig3}}
\end{figure*}

\begin{figure*}[hbt]
\centerline{\psfig{file=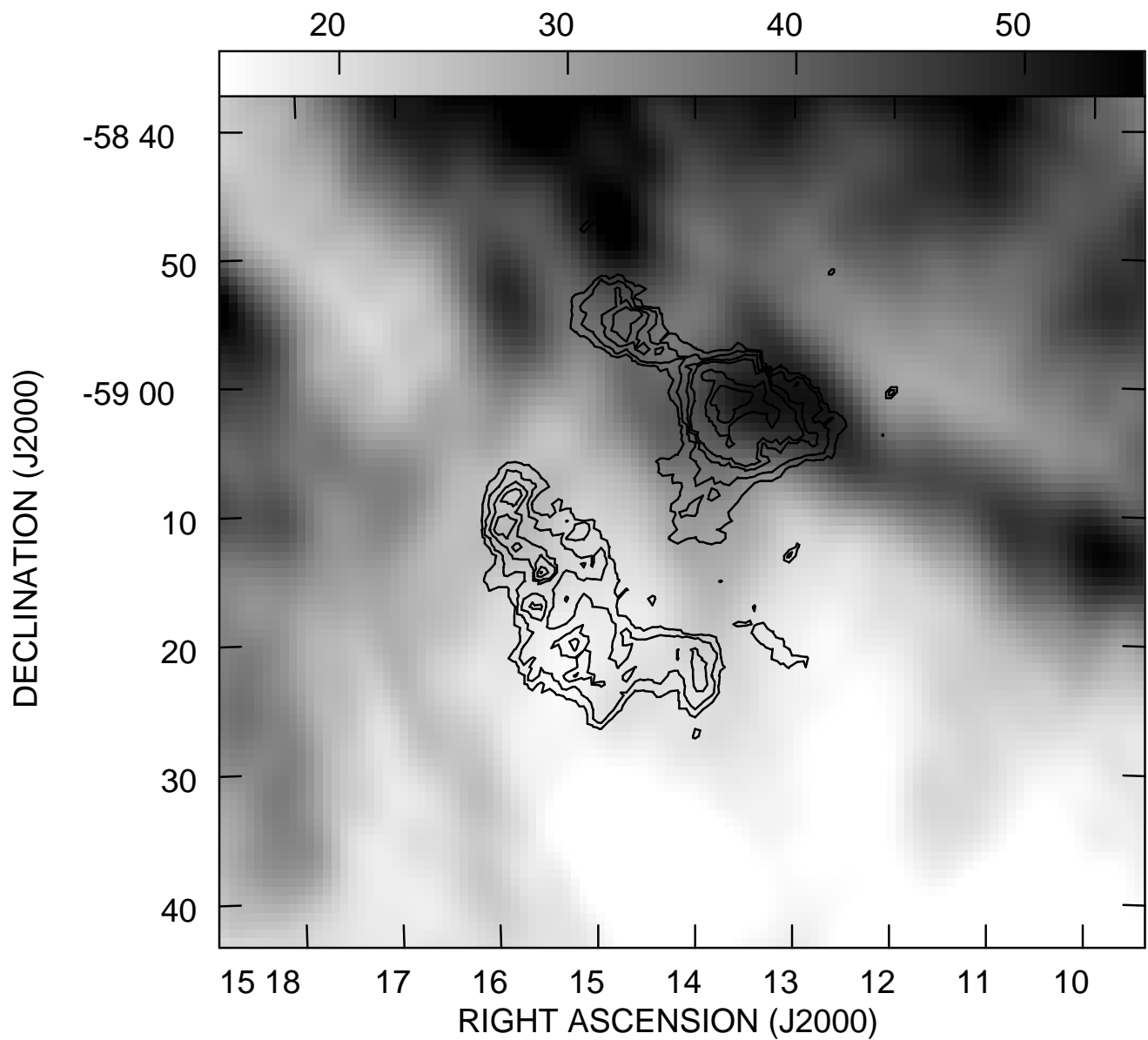,height=17cm,clip=}}
\caption{Comparison between
radio continuum and \HI\
in the velocity range --76 to --66~\kms. The grey scale, ranging from 15
to 55 mJy~beam$^{-1}$ \kms\, 
corresponds to the \HI\  distribution  
, while the contours represent the   radio continuum emission 
 at 1.4~GHz (G99.)
 Plotted contours are 5, 10, 20, 30, 60, 90 mJy~beam$^{-1}$. 
\label{fig4}}
\end{figure*}

\begin{figure*}[hbt]
\centerline{\psfig{file=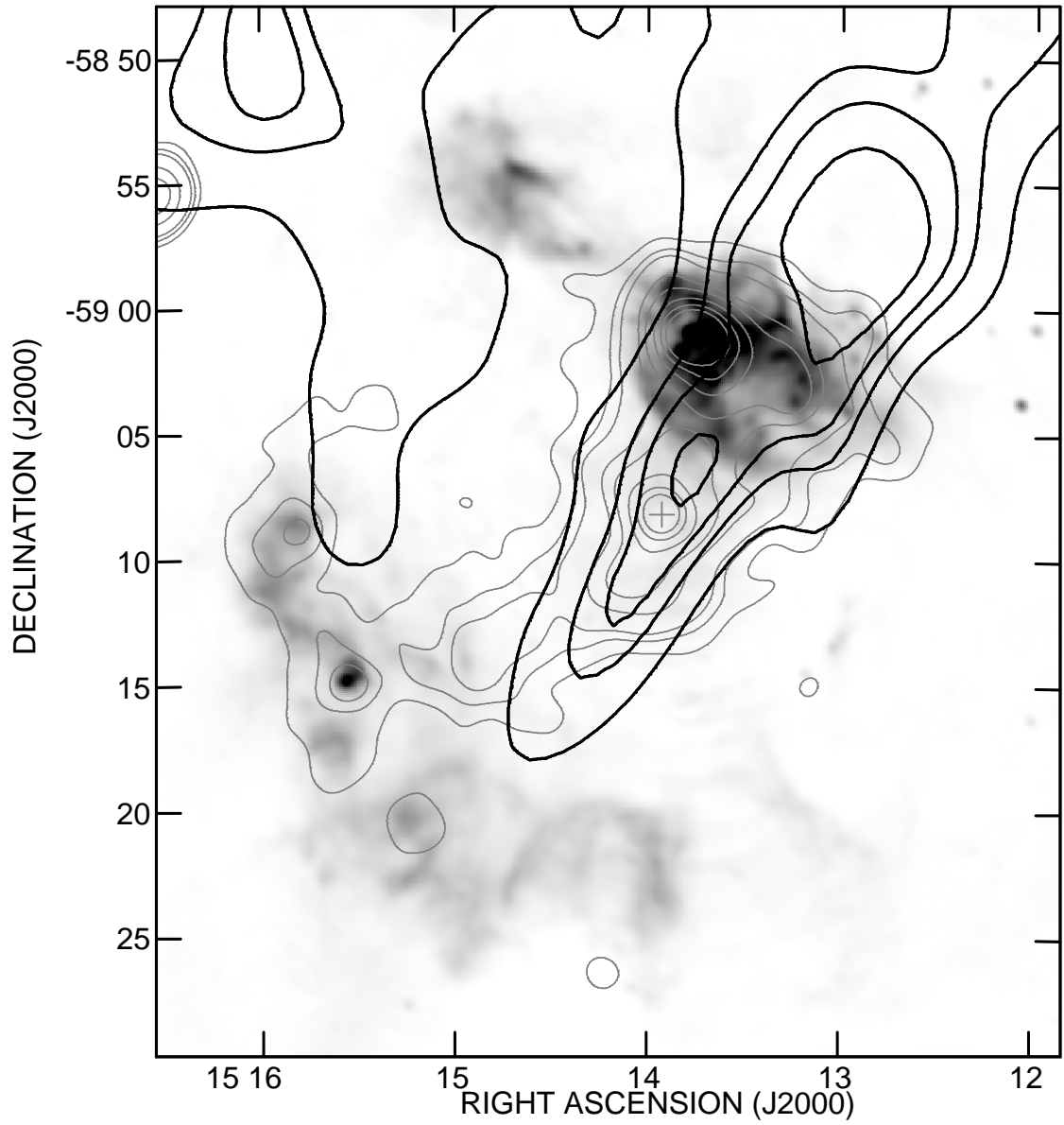,height=17cm,clip=}}
\caption{Comparison between radio continuum, X-rays and \HI\ in the
velocity range +10 and +20~\kms.
X-ray emission from the region (\cite{tmc+96}) is overlaid
in light grey contours. 
The \HI~ contours are 3, 3.6, 4.2 and 4.5 Jy~beam$^{-1}$~\ks. 
The cross indicates the position of the pulsar PSR~B1509--58.\label{fig5}}
\end{figure*}

\end{document}